%% file: MainFile.tex
\documentclass[conference]{IEEEtran}
\IEEEoverridecommandlockouts
\usepackage{cite}
\usepackage{array}   
\usepackage{algorithm}
\usepackage{algorithmic}
\usepackage{graphicx}
\usepackage{textcomp}
\usepackage[table]{xcolor}
\usepackage{enumitem}
\usepackage{tikz}
\usepackage{anyfontsize}
\usepackage{url}
\usepackage{subfig}
\usepackage{placeins}
\usepackage{soul}
\usepackage{amsmath}
\usepackage{amssymb}
\usepackage{mathptmx}  
\usepackage[top=0.7in, left=0.7in, right=0.7in, bottom=0.8in]{geometry}

\def\BibTeX{{\rm B\kern-.05em{\sc i\kern-.025em b}\kern-.08em
    T\kern-.1667em\lower.7ex\hbox{E}\kern-.125emX}}
\begin{document}

\title{Exemplifying Emerging Phishing: QR-based Browser-in-The-Browser (BiTB) Attack\\
}
\author{
    \IEEEauthorblockN{
        Muhammad Wahid Akram$^{*}$, 
        Keshav Sood$^{*}$, 
        Muneeb Ul Hassan$^{*}$, 
        and Basant Subba$^{+}$
    }
    \IEEEauthorblockA{
        \textit{$^{*}$School of IT, Deakin University, Geelong, Australia}\\
        \textit{$^{+}$Indian Institute of Technology Ropar (IIT Ropar), Punjab, India}\\
        Email: \{s224289198, keshav.sood\}@deakin.edu.au, muneebmh1@gmail.com, basant.subba@iitrpr.ac.in
    }
}

\maketitle

\begin{abstract}
Lately, cybercriminals constantly formulate productive approaches to exploit individuals. This article exemplifies an innovative attack, namely QR-based Browser-in-The-Browser (BiTB), using proficiencies of Large Language Model (LLM) i.e. Google Gemini. The presented attack is a fusion of two emerging attacks: BiTB and Quishing (QR code phishing). Our study underscores attack’s simplistic implementation utilizing malicious prompts provided to Gemini-LLM. Moreover, we presented a case study to highlight a lucrative attack method, we also performed an experiment to comprehend the attack execution on victims’ device. The findings of this work obligate the researchers’ contributions in confronting this type of phishing attempts through LLMs.
\end{abstract}


\section{Introduction and Background}
Over almost last three decades, since the inception of digitization, humans are continuously being the target of phishing scams. In cybersecurity, humans are especially considered the most vulnerable link in the success of phishing attacks~\cite{huang2022advert}. The authors in~\cite{goenka2024comprehensive} comprehensively classified various classes of mobile phishing. This includes smishing (SMS phishing), vishing (voice phishing), QRishing or Quishing (Quick Response (QR) code phishing), etc. Similarly, in~\cite{asiri2024phishingrtds} the authors proposed an innovative form of phishing attack named Browser-in-The-Browser (BiTB) attack. In fact, attackers can also launch these attacks in combination, such as sending a phishing email containing a QR code attachment that leads to a malicious website.
Furthermore, this article showcased how Large Language Models (LLMs) i.e. Google Gemini can helpful in generating malicious content for the successful execution of QR-based BiTB phishing attacks.\par
Currently, many individuals and businesses use QR codes for multiple benefits, for example, embedding website URLs in it, sending via emails, and more. Similarly, LLMs have emerged as a significant tool across several applications including AI-driven chatbots, research, innovation, and content generation to data analysis. However, their availability and productiveness make them vulnerable to exploit for creation of malicious content for phishing attacks~\cite{roy2024chatbots}. In the existing literature various works reported phishing attacks via QR codes (please note that phishing via QR codes is known as Quishing). In this regards, Bekavac et al.~\cite{bekavac2024qr} underlines QR code tampering approach in which, attackers either physically replace the whole QR code with a fake one or interfere with the pixels (white and black) of QR code. Moreover,~\cite{al2021secure} illuminates the effectiveness of barcode-in-barcode attack.\par
Regarding phishing with LLMs,~\cite{roy2024chatbots} and~\cite{taeb2024seeing} comprehensively studied LLMs and Vision Language Models (VLMs) models respectively. They highlighted the competencies of LLMs and VLMs in generating malicious content for specific phishing attacks including email phishing, Quishing, and BiTB. Generally, LLMs did not respond to given malicious prompts. However, if the same malicious prompt(s) submitted to the LLM in a sophisticated way as highlighted by~\cite{roy2024chatbots} then attackers can be successful in producing phishing content. Similarly, in context of BiTB,~\cite{tommasi2022browser} and~\cite{tzschoppe2023browser} attempts to present the practical implementation of this attack. Whereas the proposed methodologies from both studies are far more complex to implement in realistic settings.\par
Motivated from QR-phishing and involvement of LLMs in social engineering schemes, in our work, we are highlighting an evolving attack, i.e., BiTB (in QR context) named as QR-based BiTB phishing attack. The authors in~\cite{asiri2024phishingrtds} have described BiTB attack (in non QR-code scenarios), where a popup modal shows a real browser which contains a phony URL and phishing web page. This paper has taken BiTB phishing attack one step ahead from an attacker's perspective. We have shown the BiTB attack with malicious dynamic QR code and Gemini-LLM which renders it into a compelling attack vector. Therefore, the combination of dynamic QR code, and Gemini-LLM could empower BiTB attack to certainly bypass the standard phishing detection mechanisms.\par
Moreover, the aim of our study is not only to illuminate technique of this attack but also to emphasize its consequences on the attackers' capabilities. Through evaluating its anticipated impact, we are looking to contribute in continuing discussion on productive counter frameworks. Below are the contributions of our study:\par
\begin{itemize}
    \item We discussed a novel phishing attack namely QR-based BiTB which is a novel attack in QR-Codes platform and has never been studied before in the existing literature (in QR context). A step-by-step detailed discussion is given to better comprehend how to launch this attack. This discussion is extremely helpful from the reproducibility perspective (of our work).
    \item We utilized the Gemini-LLM in integrating QR-based BiTB phishing attack which generating the malicious content through providing malicious prompts. The malicious content further exploits while the execution of QR-based BiTB attack in real-time.
    \item  We presented a case study to demonstrate the execution phase of this attack which helps readers to understand a lucrative approach an attacker can use to successfully launch this attack. This is presented with an overall aim to further investigate novel threat models/approaches in future and to contribute in continuing discussion on productive counter frameworks applicable to mitigate this attack in wider attack scenarios. 
    \item A Proof-of-Concept (PoC) is given which shows a successful launch of this attack using a practical and simple approach. This is to show how publicly available tools and resources, attacker could utilize to commence this attack which is practical and easy to launch. 
\end{itemize}

\section{Overview of The QR-based Browser-in-The-Browser (BiTB) Attack}
In this section, firstly we described the capabilities of a dynamic QR code and then presented a case study to understand an attackers' approach to attract a potential victim. Rest of the section includes a high level attack framework, implementation setup, and execution part of proposed attack as depicted in Fig.~\ref{high_level}. 

\subsection{Capabilities of Dynamic QR codes}
\begin{figure}[htbp]
\begin{center}
\includegraphics[width=\linewidth]{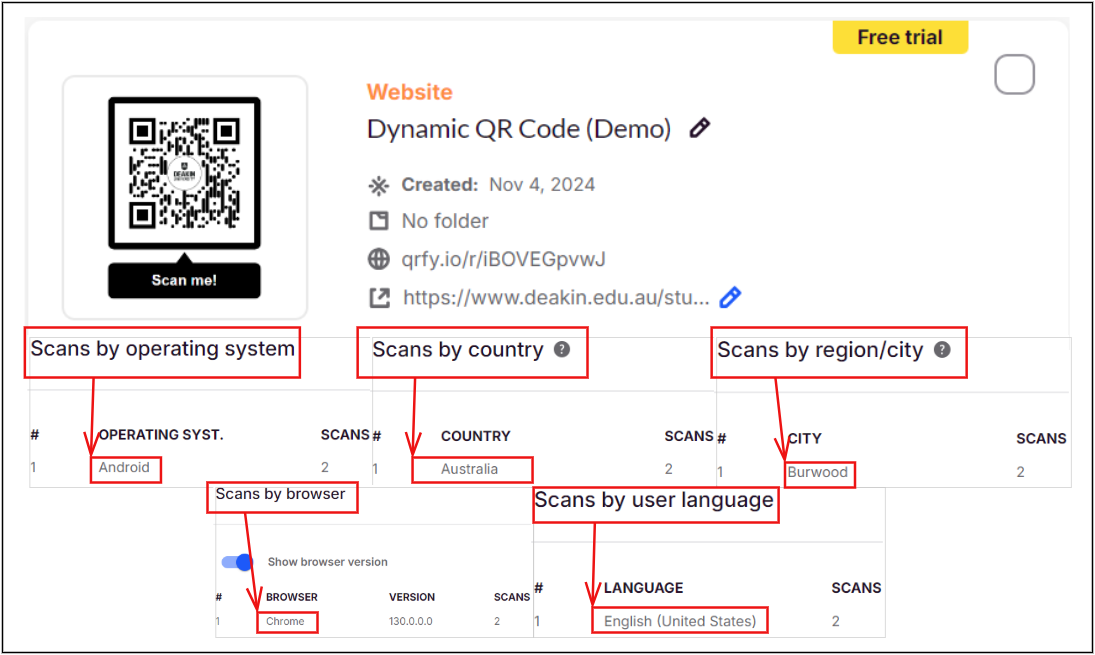}
\caption{QRFY results upon scanning dynamic QR code.}
\label{qrfy_output}
\end{center}
\end{figure}

QR codes can be of two types: static and dynamic. Static QR codes can be generated without any cost. The main difference is, when a static QR code created with some content (e.g. URL or text) it cannot be modified later. Whereas, dynamic QR codes has a feature to update the embedded content after its generation. Individuals and organizations can acquire further details whenever an individual scan a dynamic QR code such as the location, operating system (OS) of scanning device, browser type (where link opens), and current IP address of device, etc. To exemplify, we performed an experiment with an online QR code generator (which supports dynamic QR codes). We registered a trial account on QRFY\footnote{\url{https://qrfy.com/}} and generated a dynamic QR code. The results of this experiment is presented in Fig.~\ref{qrfy_output} (i). It clearly indicates that user's personal information is being sent on this site through QR code.\par

\subsection{Our Case Study}
In our experiment, we consider an attacker, who offers a potential victim the lifetime `YouTube Ads Free Access'. To redeem offer, a potential victim requires to scan the malicious dynamic QR code generated by the attacker. Upon scanning, victim redirects to the phishing site, where attacker mimicked the official web page of `YouTube premium' access i.e. \textit{https://www.youtube.com/premium} with additional BiTB attack settings which are real-time begin generating by providing specific malicious prompts to Gemini-LLM using its API. The moment malicious site opens on victims' browser, the popup modal 1 generates using Gemini-LLM and shows up as shown in Fig.~\ref{step_2} (a). This modal requests victim to input his/her `First Name'. On submitting `First Name', the popup modal 2 generates using Gemini-LLM and appears with the Google signin form Fig.~\ref{step_2} (b). When victim provides his/her signin credentials, then malicious script redirects victim to the official app/site (app in case of mobile device and site in case of desktop) of `YouTube'. Eventually, victims' credentials gets stored into the attackers' database. The detailed description of Fig. \ref{step_2} is discussed in the following sections.\par

\subsection{The High-Level Framework of the Proposed Attack}
The step wise description of the high level framework in Fig.~\ref{high_level} is discussed below:

\begin{figure}[htbp]
\begin{center}
\includegraphics[width=\linewidth]{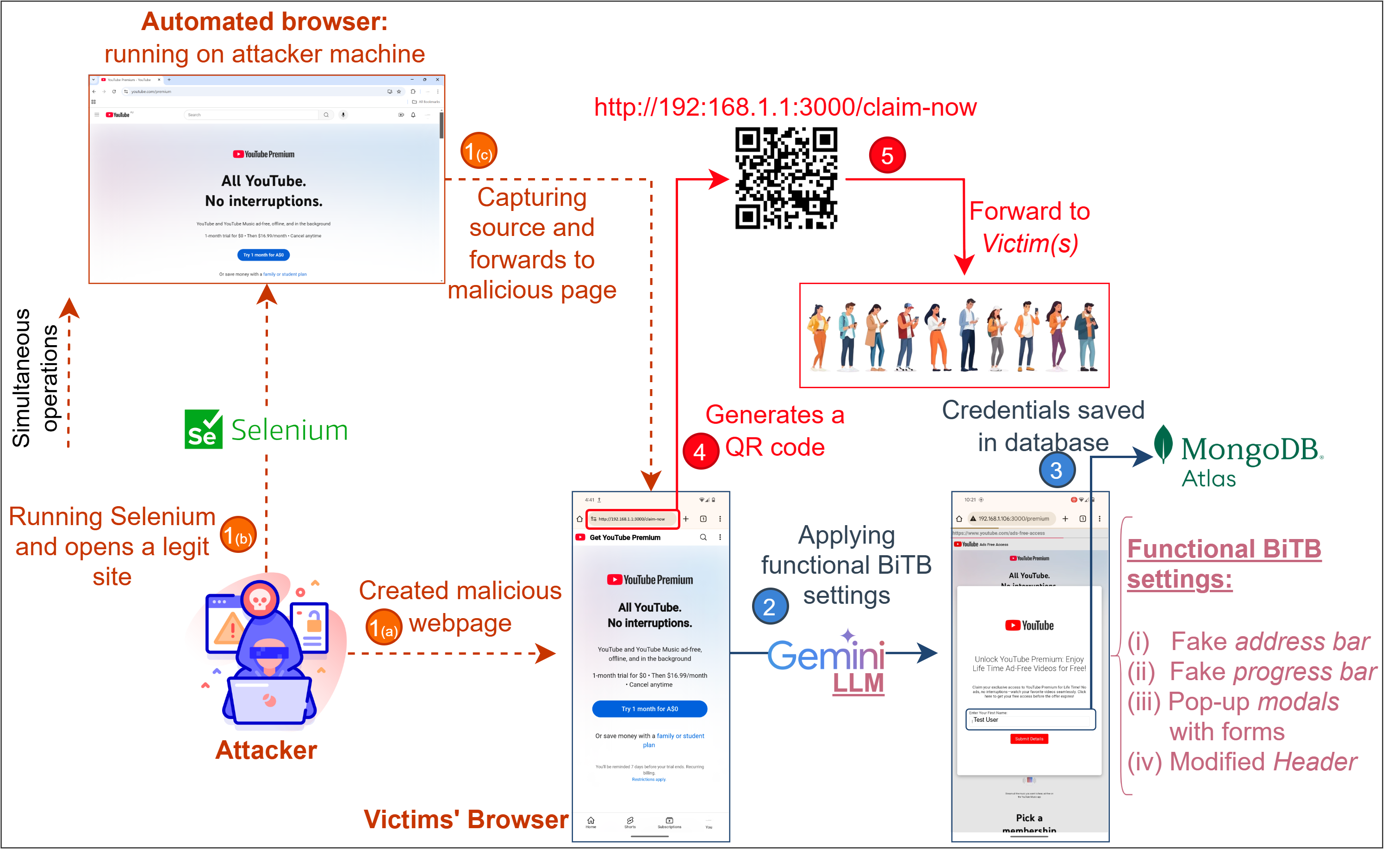}
\caption{The point-by-point flow of QR-based BiTB attack.}
\label{high_level}
\end{center}
\end{figure}

\begin{enumerate}[label=\textbf{Step \arabic*:}, leftmargin=*, align=left]
    \item The operations 1(a), 1(b), and 1(c) executes simultaneously. In 1(a), an attacker creates a malicious web page that shows up in victims' browser (Chrome in our experiment) running on local server link (i.e. \textit{http://192.168.1.1:3000/claim-now}). Afterwards, 1(b), opens the automated browser with a legit site on attackers' machine. During 1(c), the source of legit site is captured and displayed on the malicious web page (see Algorithm \ref{algo1}) using Selenium.
    \item In operation 2, attacker modifies the extracted source by applying the BiTB settings on the malicious web page using API of Gemini-LLM. This includes, appending \textit{fake address bar} on top representing legitimate URL, a \textit{fake progress bar}, a \textit{fake popup modals}, modified \textit{header}, which are generated through Gemini-LLM API. The intention of modals is to prevent victims' interaction with the web page (like scrolling) and also to collect his/her sensitive credentials.
    \item If victim falls into the trap of BiTB settings of attacker and submits his/her credentials. Then in operation 3, these credentials saved to MongoDB database.
    \item In operation 4, the attacker embeds the server link into the QR code. So that, victim cannot recognize the malicious link before accessing it.
    \item Now, attacker is ready to disperse QR-based BiTB attack and can utilize any resource to forward malicious QR code to the victims such as by sending an email with fake offer like our case study or posting on social platforms, etc.
\end{enumerate}

\section{The Attack Implementation Setup: Attackers' Side}
This section discussed the integral components involved in our experiment to launch the QR-based Browser-in-The-Browser (BiTB) attack which is divided into three modules. Two of those modules are illustrated in this section. In addition, Algorithm \ref{algo1} is also presented to better comprehend the step-by-step process of QR-based BiTB attack from attackers' end. Whereas, the third module is explained in the next section.\par

\begin{figure}[htbp]
\begin{center}
\includegraphics[width=3in, height=2.5in]{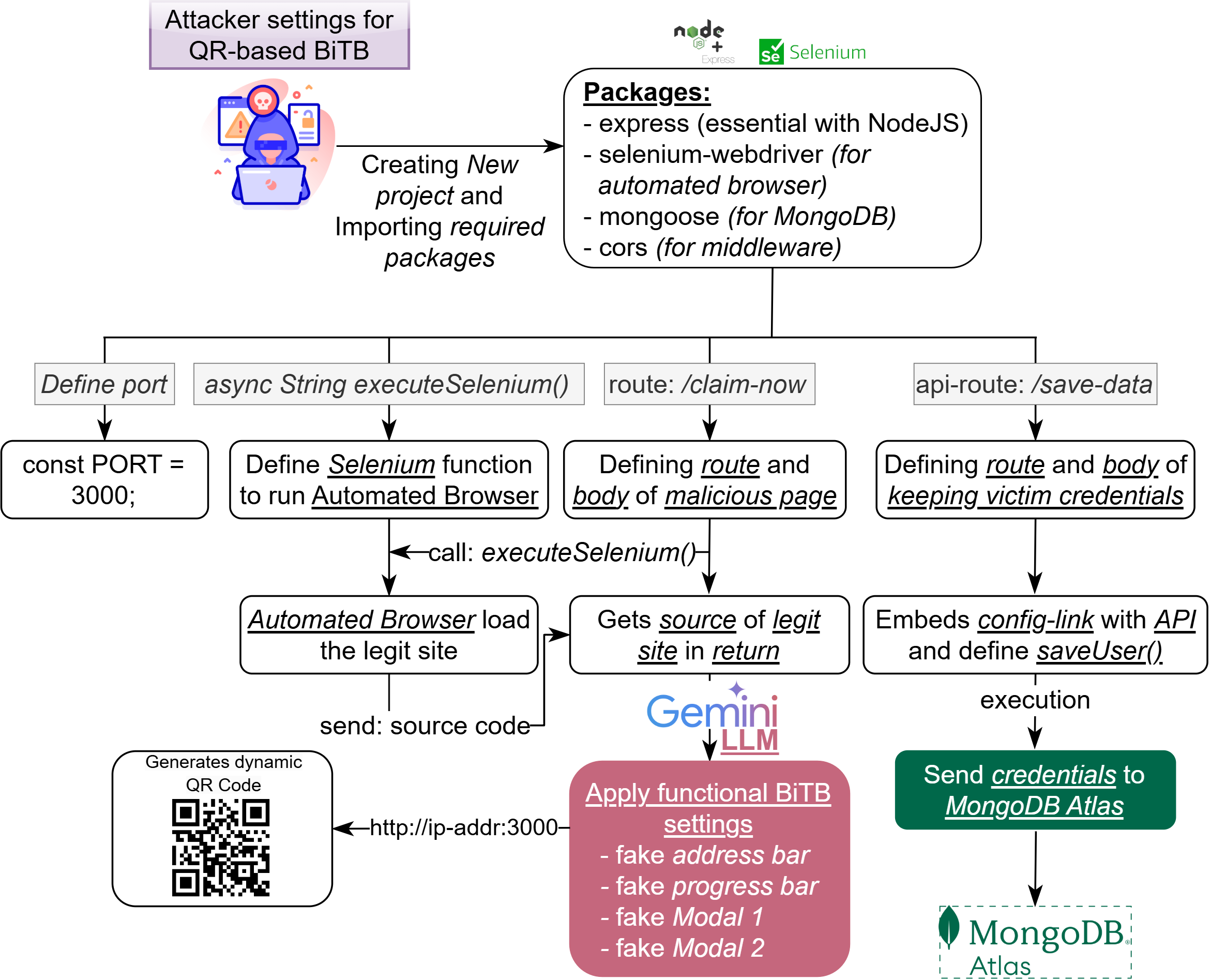}
\caption{Detailed architecture of attack implementation setup.}
\label{selenium_module}
\end{center}
\end{figure}

\subsection{Module 1:- Writing and launching the BiTB Attack using
Selenium:}
In this module, the entire setup of Selenium with NodeJS including the Gemini-LLM's API is described and pictured in Fig.~\ref{selenium_module}.

\subsubsection{New Project in NodeJS} Once NodeJS is installed, the attacker creates a \textit{New Project} with a server file \textit{bitb\_testing.js} in a specified directory. The entire malicious code is written in NodeJS including the script of Selenium, database and Gemini-LLM's APIs to design settings of BiTB.
\subsubsection{Defining PORT} To run project on local machine, a specific port is defined for example, \textit{port = 3000} as shown in Fig. \ref{selenium_module} and Algorithm \ref{algo1}.
\subsubsection{Selenium WebDriver} Next, from Algorithm \ref{algo1}, a function \textit{executeSelenium()} for Selenium script is defined which generates an automated browser (i.e. copy of Chrome) and opens a legitimate site as shown in in simultaneous operations 1(a), 1(b), 1(c) in Fig.~\ref{high_level}.

\subsubsection{Route for Malicious Web Page} Now, attacker creates a route \textit{`/clain-now'} for the malicious web page which appears on victims' browser and runs the malicious script. Firstly, the \textit{executeSelenium()} executes from its body and returns the source of legitimate site. This process is outlined in operation 2 of Fig.~\ref{high_level} and Algorithm~\ref{algo1} where Gemini-LLM is also employed to generate malicious code for BiTB settings. Later, the attacker embeds this route with malicious link in the QR code.

\subsubsection{Route for Database API} The attacker also establishes an API for database to save credentials of victim. For this, attacker defines a post request as showing in Algorithm \ref{algo1} and connect the API with online database \textit{(MongoDB Atlas)} using a \textit{config-link}.

\begin{figure}[htbp]
\begin{center}
\includegraphics[width=3.00in, height=1.00in]{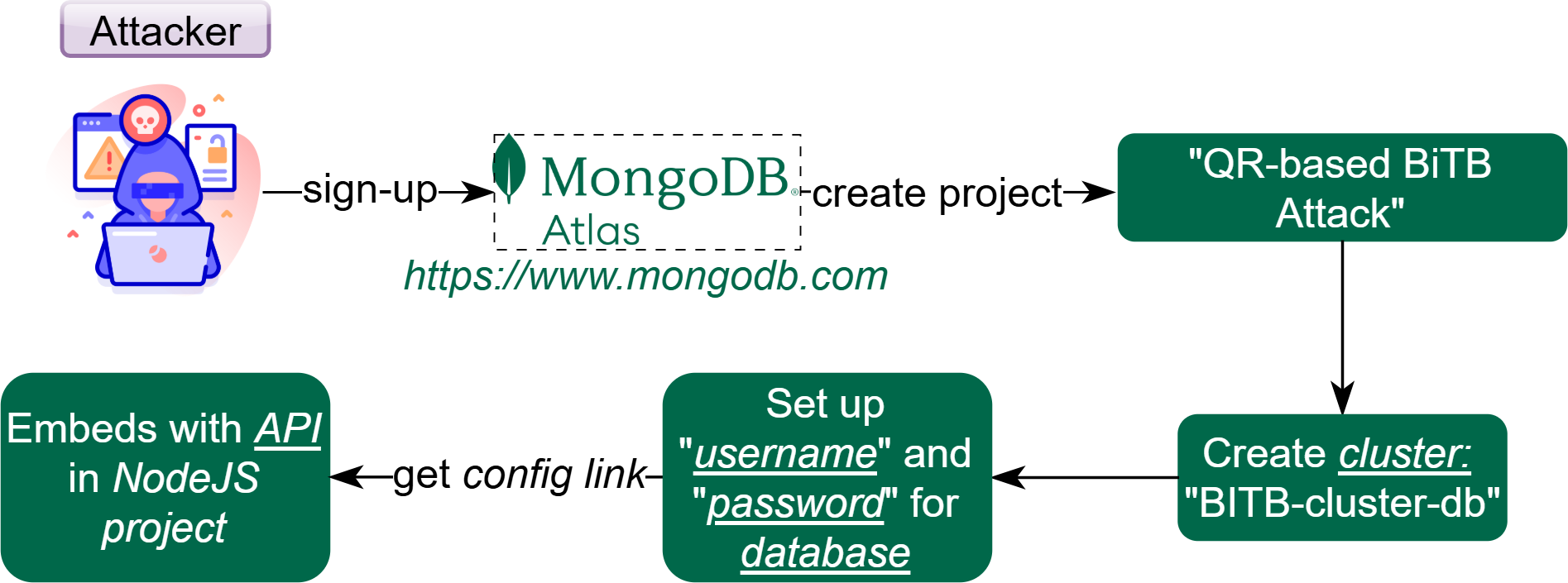}
\caption{MongoDb Atlas Setup.}
\label{mongodb_module}
\end{center}
\end{figure}

\subsection{Module 2:- Storing victim's sensitive credentials using MongoDB Atlas online database:}
Fig.~\ref{mongodb_module} illustrates the process of the attacker creating online database with MongoDB Atlas and connects this database with NodeJS project with a configuration link and API. The fundamental parts of this process are described below:

\subsubsection{MongoDB Atlas} First, attacker signed up on official site of MongoDB Atlas \textit{(https://www.mongodb.com/)} and creates a new project with name \textit{``QR-based BiTB Attack"}.
\subsubsection{Cluster and Database} To generate database inside the project, attacker creates a cluster with name \textit{``BiTB-cluster-db"} and set up the \textit{username} and \textit{password} credentials to access database.
\subsubsection{Config Link} Following this, the created cluster provides a configuration link and attacker modifies this link with credentials to use in NodeJS project for database connectivity.

\begin{algorithm}
\caption{Attackers' settings for QR-based BiTB Attack}
\label{algo1}
\scriptsize
\begin{algorithmic}[1]
\STATE \texttt{\textbf{import} libraries and set-up \textbf{PORT = 3000}}

\STATE \vspace{0.1cm} \texttt{\textbf{function} executeSelenium()}
\STATE \quad \texttt{pageSource} $\gets$ \texttt{callAutomatedBrowser(`https://youtube.com')}
\STATE \quad \texttt{\textbf{return} pageSource}

\STATE \vspace{0.1cm} \texttt{\textbf{app.get}(`/claim-now')}
\STATE \hspace*{0.1em} \texttt{modifiedContent} $\gets$ \texttt{executeSelenium()}
\STATE \quad \texttt{modifiedContent} $\gets$ \texttt{pageContent} + \texttt{GemeniLLM\_API(`malicious prompts')}
\STATE \quad \texttt{\textbf{send}(modifiedContent)}

\STATE \vspace{0.1cm} \texttt{\textbf{app.post}(`/api/save-user')}
\STATE \quad \texttt{save\_result} $\gets$ \texttt{\textbf{saveUser}(first\_name, email, password)}
\STATE \quad \texttt{\textbf{return} `User saved successfully'}

\STATE \vspace{0.1cm} \COMMENT{\textit{Gemini-LLM generate and operates below function}}
\STATE \texttt{\textbf{function} saveUser(first\_name, email, password)}
\STATE \quad \texttt{user} $\gets$ \texttt{new User(\{first\_name, email, password\})}
\STATE \quad \texttt{\textbf{return} `Success'}

\STATE \vspace{0.1cm} \COMMENT{\textit{Gemini-LLM generate and operates below functions}}
\STATE \vspace{0.05cm} \texttt{\textbf{function} updateFakeAddressBar()}
\STATE \quad Input new legitimate address into the address bar

\STATE \vspace{0.05cm} \texttt{\textbf{function} runFakeProgressBar()}
\STATE \quad Progress bar executes

\STATE \vspace{0.05cm} \texttt{\textbf{function} modifyHeader()}
\STATE \quad Changes applied to header

\end{algorithmic}
\end{algorithm}

\section{QR-based Browser-in-The-Browser (BiTB) Attack Execution: Victims' side}
This section presents the execution of QR-based BiTB attack from victims' end. Moreover, the working of first (selenium module) and second (mongodb atlas module) module can be seen in this (third) module. Additionally, Algorithm \ref{algo2} is scripted to better comprehend the method.\par

\begin{figure}[htbp]
\begin{center}
\includegraphics[width=3.0in, height=2.8in]{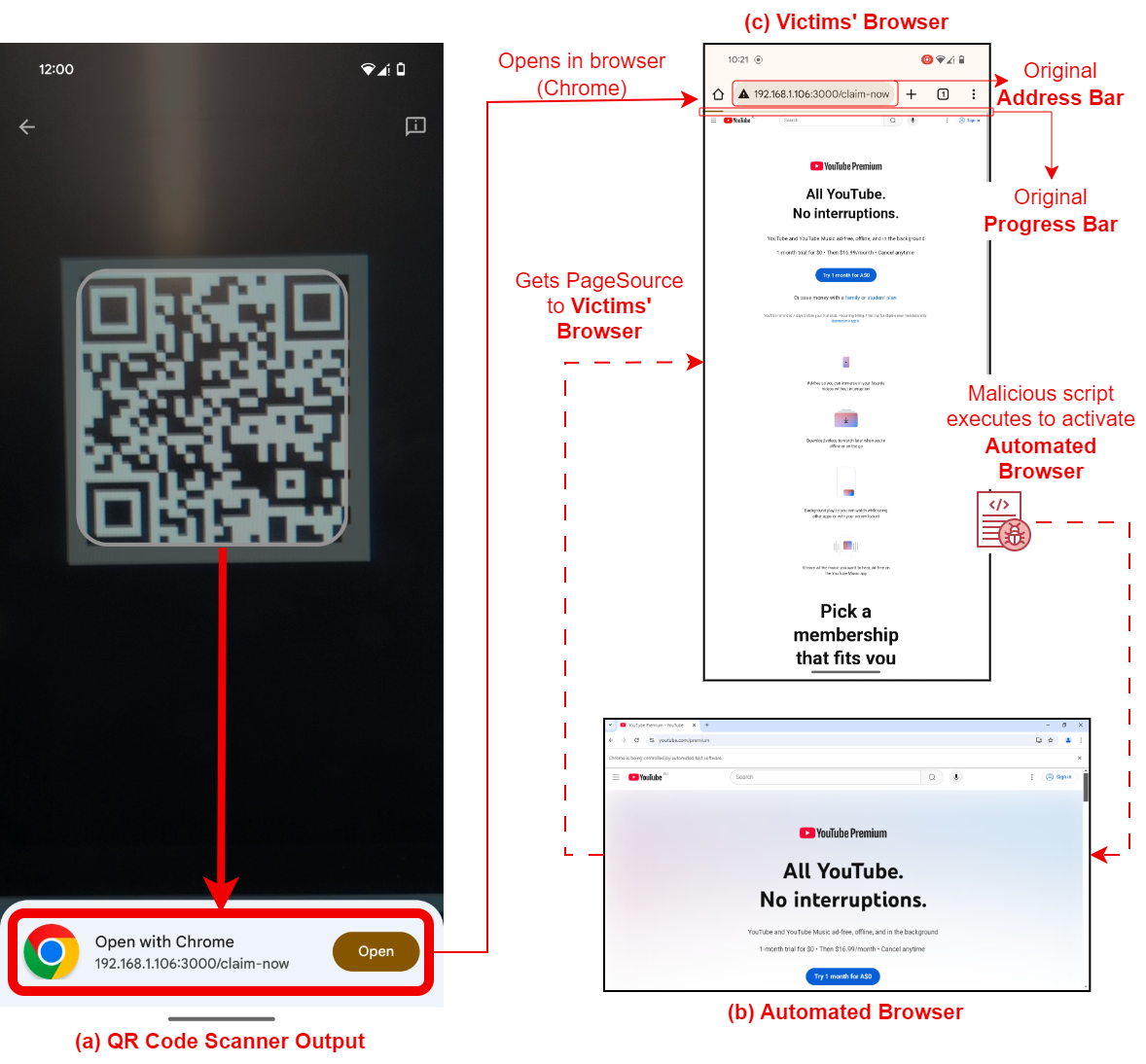}
\caption{(a) QR Code Scanner output, (b) Automated Browser, and (c) Victims' Browser.}
\label{step_1}
\end{center}
\end{figure}

For execution of BiTB attack, we deploy a Windows laptop where module 1 and module 2 are already configured. The NodeJS project is executed on local server url: \textit{`http://localhost:3000'}. By default, this URL is inaccessible on mobile. To enable this, attacker uses current IP address of machine which is connected with available Wi-Fi network. Afterwards, the mobile accessible URL embeds in malicious QR code was\textit{`http://ip-addr:3000'}. In Fig.~\ref{step_1} (a) and Algorithm~\ref{algo2}, victim scans the malicious QR code using built-in QR scanner of mobile device. As a result, he/she gets embedded link to open in default browser. The moment link open in browser, the malicious script executes and initiates the \textit{Automated Browser} on attackers' machine with a legitimate link of \textit{https://www.youtube.com/premium} as scripted in Algorithm~\ref{algo2}. The malicious script waits until page loads in \textit{Automated Browser} and instantly it returns the source code to the victims' browser as shown in Fig.~\ref{step_1} (b) and (c), where, the original address and progress bars are also visible.\par

\begin{figure}[htbp]
\begin{center}
\includegraphics[width=3.0in, height=2.8in]{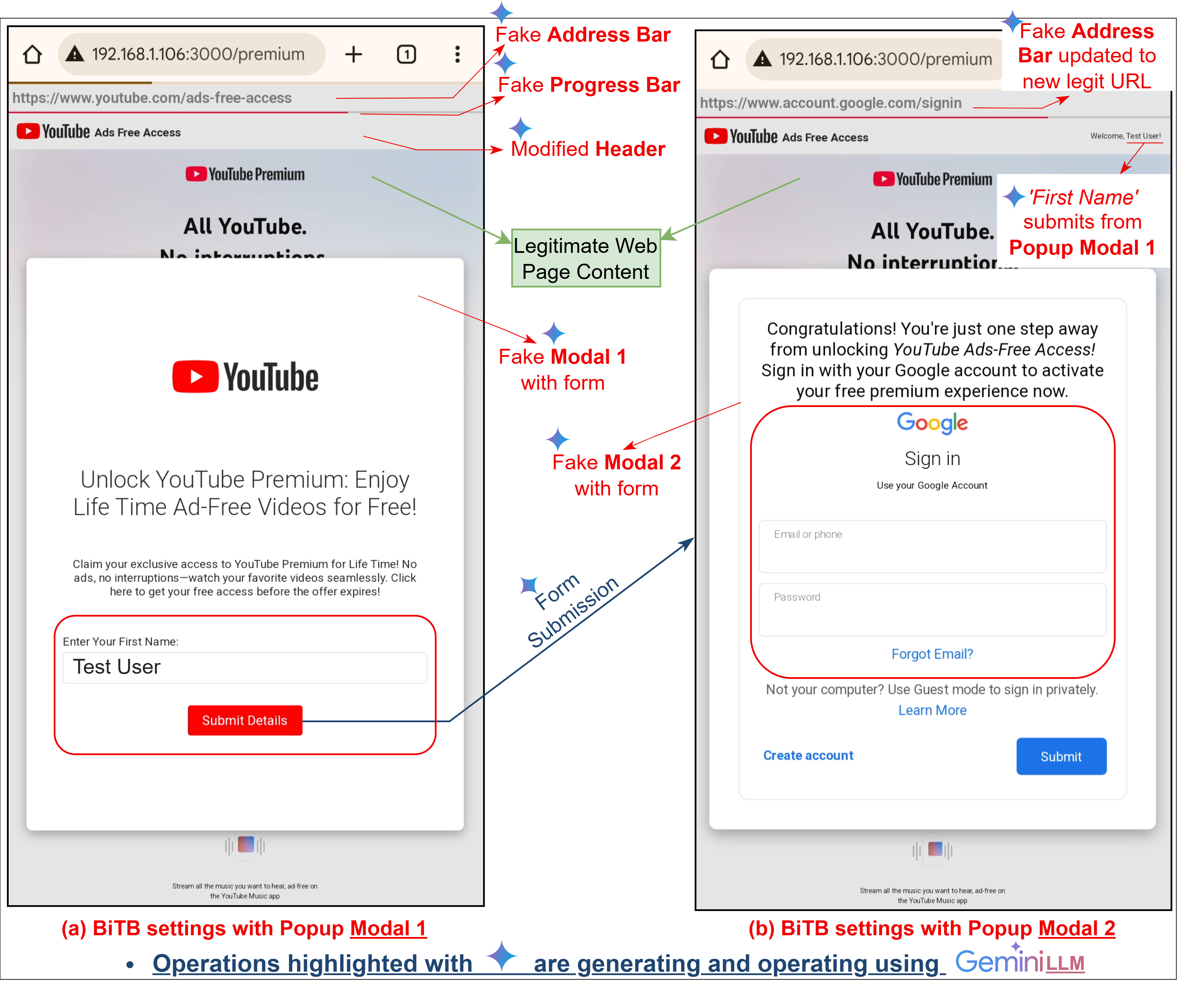}
\caption{(a) BiTB settings with Popup Modal 1 and (b) BiTB settings with Popup Modal 2.}
\label{step_2}
\end{center}
\end{figure}

While loading of the malicious web page, the requests of malicious prompts also simultaneously sent to the Gemini-LLM by triggering its API to implement and present the BiTB attack settings on the malicious web page as highlighted in Fig.~\ref{step_2} (a) and Fig.~\ref{step_2} (b). This includes, alluring heading and paragraph texts, fake address bar displaying legit URL, fake progress bar, modified header, a fake popup modal 1, and modal 2 to acquire victim details. These settings from attacker can easily trick an individual. In Fig.~\ref{step_2} (a) and Algorithm~\ref{algo2}, popup modal 1 requires victim to input his/her \textit{`First Name'} to proceed further with claiming the offer. On submission, the popup modal 1 hides and popup modal 2 shows up on the malicious page as represented in Fig.~\ref{step_2} (b). Also, header and progress bar of the page is further modified with \textit{`First Name'} of the victim using malicious script. Whereas, the URL in fake address bar is also altered to new legit URL. In addition, the attacker keeps entered \textit{`First Name'} in variable \textit{first\_name} as highlighted in Algorithm~\ref{algo2}.\par

\begin{figure}[htbp]
\begin{center}
\includegraphics[width=3.0in, height=2.8in]{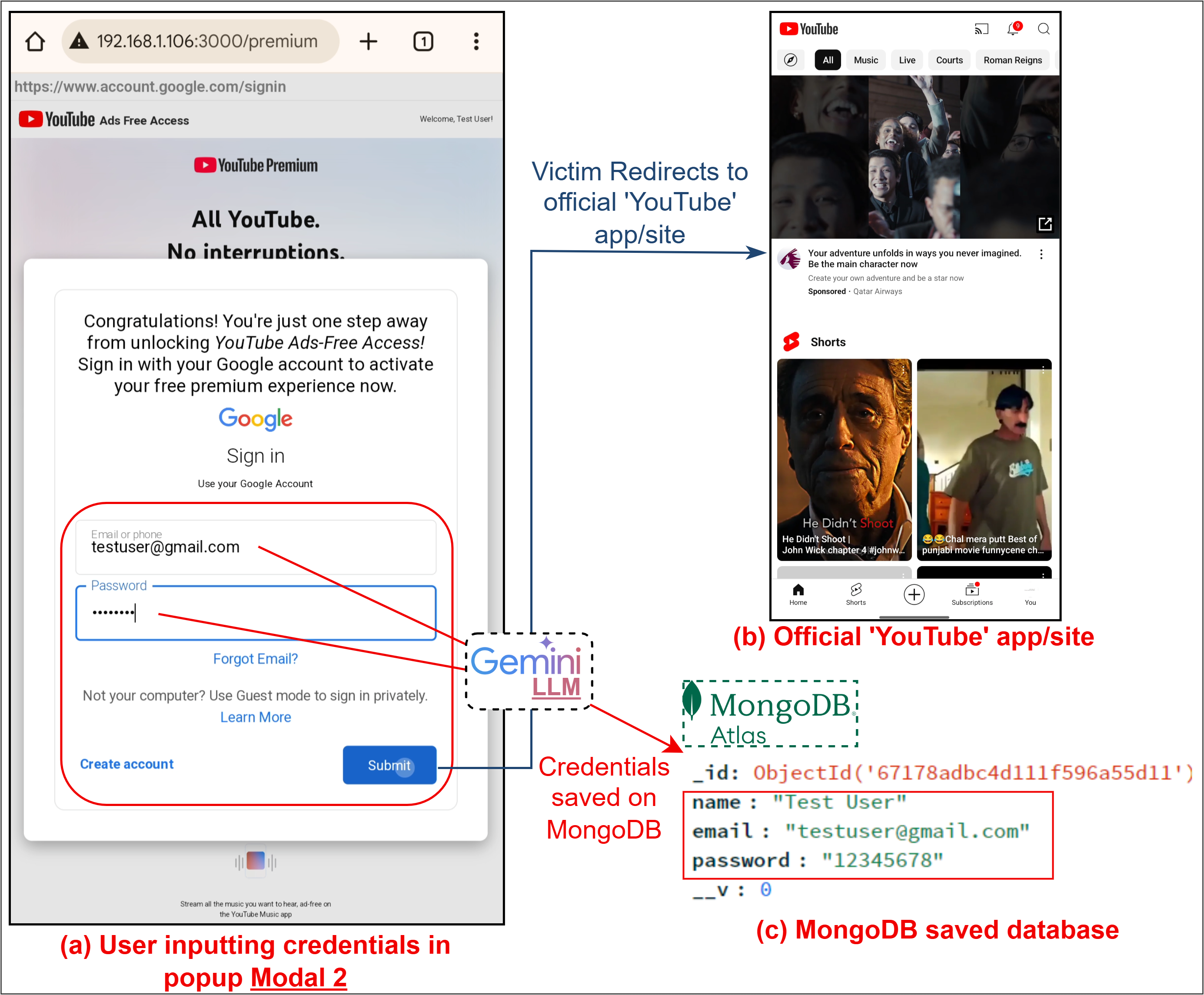}
\caption{(a) User inputting credentials in popup Modal 2, (b) User redirects to official \textit{`YouTube'} app/site, and (c) Screenshot from MongoDB database.}
\label{step_3}
\end{center}
\end{figure}

Modal 2 is the final stage of the BiTB attack, where attacker laid the trap of stealing victims' sensitive credentials. To manifest the legitimacy, attacker mimics the Google sign-in form in modal 2 as depicted in Fig.~\ref{step_3} (a) and Algorithm~\ref{algo2}, whenever victim inputs his/her credentials in fake Google form and submits it. Then after submission, the fake progress bar refreshes again. Moreover, from Fig.~\ref{step_3} (b) and Algorithm~\ref{algo2}, victim redirected to the official app of \textit{`YouTube'} if he/she is using a smartphone. For desktop users, they are redirected to official site of \textit{`YouTube'}. Simultaneously, victims' credentials gets stored in MongoDB database. In addition, dynamic QR codes captures individuals personal details upon scanning as shown in Fig. \ref{high_level}. This signifies that the attacker eventually holds the credentials of a victim using QR-based BiTB attack. Here, it is worth noticing that all of above steps are executed and generated in real-time using the malicious prompts which are requested to Gemini-LLM. This indicates the explicit illustration of how LLMs can be maliciously utilized in the execution of phishing attacks like the one proposed in this article i.e. QR-based BiTB. 

\begin{algorithm}
\caption{Execution of QR-based BiTB attack from victims' end}
\label{algo2}
\scriptsize
\begin{algorithmic}[1]
\REQUIRE Algorithm \ref{algo1} already defined, Malicious QR code embeds with a URL

\STATE \vspace{0.2cm} Call \texttt{\textbf{app.get}(`/claim-now')} from Algorithm~\ref{algo1}
\STATE Call \texttt{\textbf{function} executeSelenium()} from Algorithm~\ref{algo1}

\STATE \vspace{0.2cm} Modal 1 operates on victims' browser and require `First Name' using \textit{Gemini-LLM}
\IF{Details submitted}
    \STATE \textbf{Hide} Modal 1 and keep input value in variable \texttt{`first\_name'}
    \STATE Call \texttt{modifyHeader()} from Algorithm~\ref{algo1}
    \STATE Call \texttt{updateFakeAddressBar()} from Algorithm~\ref{algo1}
    \STATE Call \texttt{runFakeProgressBar()} from Algorithm~\ref{algo1}
    \STATE \textbf{Reveal} Modal 2
\ENDIF

\STATE \vspace{0.2cm} Modal 2 operates on victims' browser and require `Email' and `Password' using \textit{Gemin-LLM}
\STATE Victim inputs credentials and submits it
\IF{Details submitted}
    \STATE Call \texttt{\textbf{runFakeProgressBar}()} from Algorithm~\ref{algo1}
    \STATE \textbf{Hide} Modal 2 and keep input values in variable \texttt{`email, password'}
    \STATE \vspace{0.1cm} Call \texttt{\textbf{app.post}(`/api/save-ser')} from Algorithm~\ref{algo1}
    \STATE \textbf{Redirects} victim to official app/site
\ENDIF

\end{algorithmic}
\end{algorithm}

\subsection{Key Insights}
To sum up, the execution of QR-based BiTB attack from scanning malicious QR code to showcasing multiple BiTB settings, the attacker drives victim(s) during the whole process of phishing attempt. Below are the core findings of our work:
\begin{itemize}
    \item In our experiments, the manipulation of fake configurations are much simple and effective in terms of vulnerabilities. We depicted how attacker utilizes easily accessible tools to execute the proposed phishing activity.
    \item QR-based BiTB using Gemini-LLM brought up as a novel social engineering attack technique that emphasizes essential demand to further strengthen defenses that can effectively withstand the evolving attack challenges.
    \item We demonstrated our case study using Chrome browser on Android mobile, however, because of attacks' generalizability initiated from scanning a QR code to a browser where subsequent steps executes, this attack is practically possible with any browser and smartphone device.
\end{itemize}

\section{Conclusion and Future Directions}
Our work shows a peculiar form of a novel phishing attack i.e., QR-based BiTB attack using Gemini-LLM. The impact of this attack magnified due to the concatenation of two innovative attacks including Quishing and BiTB, and utilization of LLM. We demonstrated how simple compositions of open-source available resources can initiate this attack. Attackers continuously developing contemporary tactics to mislead users through presenting seemingly authentic information. The proposed QR-based BiTB is a prime example of it.\par

Lastly, this study has few shortcomings. Firstly, we utilized Selenium webdriver tool on local server, however, this tool is not supported by online servers due to its appearance in GUI form like an actual browser (i.e. Chrome). Secondly, operating this attack on local server requires attackers' machine and victims devices to connected on same network. This is because an attacker needs to use the IP address while connecting with network. We aim to address these gaps in future.\par

\input{./References/MainFile.bbl}

\end{document}